\documentclass[citeautoscript,floatfix,twocolumn]{revtex4-1}
\usepackage[decmulti]{inputenc}

 \usepackage{amsmath}
 \usepackage{amsfonts}
 \everymath{\displaystyle}

\usepackage{graphicx}
\usepackage{epsfig}
\usepackage{epstopdf}
       \usepackage{ulem,color,amsmath}
\usepackage[version=3]{mhchem} % Formula subscripts
\usepackage{hyperref}
\hypersetup{
  colorlinks=true,
  linkcolor=blue,
  citecolor=red,
  linktoc=all,
  pdftitle=Short range DFT combined with long-range local RPA within a range-separated hybrid DFT framework,
  pdfdisplaydoctitle=true,
  pdfpagelayout=SinglePage,
  pdfstartview=Fit,
  pdfstartpage=1,
  bookmarksopen=false
}
\newcommand{\bild}[3]{
\begin{figure}[!htb]
  \begin{center}
    \includegraphics[width=\linewidth]{#1.eps}\quad
  \end{center}
    \caption{#3}
    \label{#1}
\end{figure}
}
\begin{document}

%\begin{frontmatter}

\title{Short range DFT combined with long-range local RPA within a range-separated hybrid DFT framework} 

\author{E.\ Chermak}
\affiliation{Laboratoire de Chimie Th\'eorique, Universit\'e{} Pierre et Marie Curie\\ 4 place Jussieu, F --- 75252 Paris, France}
\affiliation{CNRS, UMR 7616, 4 place Jussieu, F --- 75252 Paris, France}
\author{ B.\ Mussard}
\affiliation{Universit\'e{} de Lorraine, CRM2, UMR 7036, Vand{\oe}uvre-l\`es-Nancy, F --- 54506, France}  
\affiliation{CNRS, CRM2, UMR 7036, Vand{\oe}uvre-l\`es-Nancy, F --- 54506, France}  
\author{J.G.\ \'Angy\'an}
\email{Janos.Angyan@univ-lorraine.fr} 
\affiliation{CNRS, CRM2, UMR 7036, Vand{\oe}uvre-l\`es-Nancy, F --- 54506, France}  
\affiliation{Universit\'e{} de Lorraine, CRM2, UMR 7036, Vand{\oe}uvre-l\`es-Nancy, F --- 54506, France}  
\affiliation{``CEU IAS'' (Central European University, Institute of Advanced studies) N\'ador utca 13, Budapest 1051, Hungary}
\author{P.\ Reinhardt}
\email{Peter.Reinhardt@upmc.fr}
\affiliation{Laboratoire de Chimie Th\'eorique, Universit\'e{} Pierre et Marie Curie\\ 4 place Jussieu, F --- 75252 Paris, France}
\affiliation{CNRS, UMR 7616, 4 place Jussieu, F --- 75252 Paris, France}

%%%%%%%%%%%%%%%%%%%%%%%%%%%%%%%%%%%%%%%%%%%%%%%%%%%%%%%%%%%%%%%%
\begin{abstract}
Selecting excitations in localized orbitals to calculate long-range 
correlation contributions to range-separated density-functional theory can
reduce the overall computational effort significantly. Beyond simple selection
schemes of excited determinants, the dispersion-only approximation, which
avoids counterpoise-corrected monomer calculations, is shown to be
particularly interesting in this context, which we apply to the random-phase
approximation. The approach has been tested on dimers of formamide, water,
methane and benzene.  
\end{abstract}
%\begin{keyword}
%range separated hybrid \sep DFT \sep random phase approximation (RPA)  \sep
%London dispersion forces \sep local correlation methods \sep localized
%orbitals \sep 
%\end{keyword}

%\end{frontmatter}
\maketitle

\section*{Introduction}
As it is now widely recognized, standard functionals fail dramatically when
calculating the interaction energy of rare-gas
dimers~\cite{patton_application_1997,Kristyan:94}. Local-density
approximations (LDA) overbind heavily~\cite{harris_simplified_1985} while
hybrid functionals produce almost all possible types of results including lack
of binding, like in the case of the popular B3LYP functional, to relatively
reasonable van der Waals minima, and sometimes strong
overbinding~\cite{Gerber:07}. The imprecisions in describing London
dispersion-type correlation effects have far-reaching implications, which are
not limited to the flaws in modeling weakly bound van der Waals complexes. For
instance, functionals that are popular in modern quantum chemistry due to
their ability to describe thermochemistry data with a reasonable
accuracy~\cite{koch_guide_2001}, while keeping the computational requirements
on a low level as compared to wavefunction-based calculations, may lead to
systematic errors in some specific cases, like the isodesmic stabilization
energies of $n$-alkanes~\cite{wodrich_systematic_2006}. Therefore we could
witness a rapidly increasing interest in new methodologies aiming at
completing well-known density functionals by adding systematically the missing
dispersion interactions. This may be achieved, for instance, by adding
non-local density functional corrections based on some fundamental
considerations about the LDA response function of an inhomogeneous electron
gas~\cite{Langreth_1996,Langreth_2010}. A much more pragmatic
way is to add atom-atom corrections, parameterized either
semi-empirically~\cite{grimme_accurate_2004,Grimme:10}, or by using more
general considerations, like in the case of the so-called exchange-hole
model~\cite{becke_exchange-hole_2005} or in the
Tkatchenko-Scheffler method~\cite{Tkatchenko:09}. A considerably more
demanding computational approach is to add missing explicit correlation
contributions to the total energy obtained from Kohn-Sham orbitals via the
adiabatic connection fluctuation-dissipation theory (ACFDT) scheme, using the
random phase approximation (RPA). The main advantage of these methods is that
while the local and semi-local correlation functionals miss long-range
correlation effects, the methods based on the RPA provide an accurate
estimation for these. On the other hand, RPA is significantly worse in
reproducing short-range correlation effects, for which standard functionals
perform better.
  
Range-separation is one of the modern tools to overcome the deficiencies of
density-functional methods in dealing with long-range dispersion
interactions. It is based on a single parameter connecting pure DFT (Kohn-Sham
calculations) to Hartree-Fock and post-Hartree-Fock
approaches~\cite{Seminario} by an appropriate non-linear scaling of the electron-electron
interactions. The short-range interaction part is taken into account by
density-functional theory, while long-range exchange and correlation
contributions are left to an ab initio wavefunction-based treatment. The
Hartree-Fock type description of the long-range exchange turns out to be a
generalization of the concept of hybrid density functionals, sometimes evoked
as range separated hybrid (RSH) method, which produces independent-particle,
single-determinant wave functions. From a density functional viewpoint such an
approach is a Generalized Kohn-Sham (GKS) scheme, which is an appropriate
starting point for a wave function treatment of long-range electron
correlation by variational or perturbational techniques. 

In view of improving DFT methods for dealing with van der Waals complexes
various range-separation based correlation approaches have been proposed,
ranging from second-order perturbation theory~\cite{angyan_van_2005} to most
elaborate coupled-cluster approaches~\cite{goll_short-range_2006}. An
intermediate 
ab-initio correlation method, the random-phase approximation, which sums
certain perturbation diagrams to infinite order, has also been adapted to the
range-separated
scheme\cite{Toulouse:09,janesko_long-range-corrected_2009,zhu_range-separated_2010,toulouse_closed-shell_2011,Angyan:11}
and has lead to good results. 

Since in the London dispersion interaction problem we are mainly concerned
with long-range electron correlations, the selection of the most relevant
excited determinants can be enormously improved by working in a localized
one-electron basis, i.e.\ in localized orbitals. This philosophy, which
follows the pioneering works by Kapuy~\cite{Kapuy:91} and by Pulay~\cite{Pulay_localizability_1983}, is the
basis of the 
family of local correlation methods, where the selection of the relevant
excitation is usually done after criteria of spatial distance between the
centroids of localized
orbitals~\cite{schutz_local_1998,BenAmor_direct_2011}. Beyond a substantial
speed-up, local correlation schemes are able to exclude a priori the
excitations that contribute to the basis set 
superposition effects (BSSE). 

The purpose of the
present work is to explore the possible advantages of a local RPA method for
the calculation of dispersion energies. We recall in a first section the basic
ingredients of RSH scheme: the 
short-range DFT and the long-range RPA, as well as the two practical aspects
which consist in the construction of localized orbitals and the selection of
the relevant excitations. In a second part, section~\ref{sec:results}, results
on some selected systems (dimers of water, methane, formamide and benzene) are
presented and discussed, and the significant simplifications achieved via the
dispersion-only approximation will be summarized in our conclusions. Some
technical details are collected in the Appendix.

\section{Theory}
\subsection{Range-separated density-functional theory}

The electron-electron (e-e) interaction operator of the electronic Hamiltonian
can be split rigorously as a combination of a long-range (lr) contribution,
which dominates almost exclusively the interaction from a given e-e distance,
and a complementary short-range (sr) contribution, which has a Coulomb
singularity when the interelectronic distance approaches to zero. Such a
separation can be achieved in several alternative ways, e.g.\ by using the
error function splitting: 

\begin{equation}
 \frac{1}{r_{ij}}\ \ \ = w_{ee}^{\text{lr}} +   
            \left(\frac{1}{r_{ij}}-
            w_{ee}^{\text{lr}} \right) \quad \text{with} \quad w_{ee}^{\text{lr}}
          (r_{ij}) \ = \ \frac{\text{erf}(\mu\, r_{ij})}{r_{ij}}
\label{eqn:rs}
\end{equation}
The parameter $\mu$ (more precisely its inverse) governs the range separation,
i.e.\ it is proportional to the distance where the sr-contribution becomes
negligible besides the lr one. 

The, in principle exact, ground state energy of a many-electron system can be
obtained in a two-step process via:  

\begin{equation}
E = E_{\text{RSH}} + E_c^{\text{lr}} 
\label{eq:srlr}
\end{equation} 
where $E_c^{\text{lr}}$ is the long-range correlation energy, usually
approximated by some wave-function method, and $E_{\text{RSH}}$ is given by  

\begin{equation}
E_{\text{RSH}} = \underset{\Phi}{\text{min}}\left\{  
                \langle\Phi\vert\hat{T}+\hat{V}_{\text{ne}} +
                \hat{W}_{\text{ee}}^{\text{lr}}  
                \vert\Phi\rangle  
                +
                E^{\text{sr}}_{\text{Hxc}}\left[n_{\Phi}\right]  
              \right\}
\end{equation} 
with the kinetic energy operator $\hat{T}$, the nuclei-electron interaction
operator $\hat{V}_{ne}$ and the electron-electron interaction
$\hat{W}_{\text{ee}}^{\text{lr}}$ written with
$w_{ee}^{\text{lr}}$. $E^{\text{sr}}_{\text{Hxc}}\left[n_{\Phi}\right]$ is the
short-range $\mu$-dependent Hartree-exchange-correlation functional, and
$\Phi$ is a single-determinant wave function. 

The minimizing single determinant is given by the Kohn-Sham-like one-electron
equations with the full-range Hartree interaction of electrons
$\hat{V}_{\text{H}}$, the long-range part of the Hartree-Fock type exchange
operator $\hat{V}_{\text{x}}^{\text{lr}}$, and a short-range
exchange-correlation potential, $\hat{V}_{\text{xc}}^{\text{sr}}$ related to
the sr xc functional, $E^{\text{sr}}_{\text{xc}}\left[n_{\Phi}\right]$: 

\begin{equation}
\left( \hat{T} + \hat{V}_{\text{ne}} + \hat{V}_{\text{H}} +
        \hat{V}_{\text{x}}^{\text{lr}} + 
        \hat{V}_{\text{xc}}^{\text{sr}} \right)\,|\phi^{\text{RSH}}\rangle\ =
        \ \epsilon\, |\phi^{\text{RSH}}\rangle
\label{eq:KohnSham}
\end{equation} 

Since we are interested in a solution of the RSH equations
(\ref{eq:KohnSham}) in localized orbitals, the iterative
Singles-Configuration-Interaction scheme described in
Ref.~\cite{reinhardt_fragment-localized_2008} will be preferred to the
conventional iterative diagonalization of the RSH matrix. This
procedure has the advantage of leaving the final orbitals as close as possible
to the set of starting orbitals, both for occupied and virtual orbitals, and
allows us to maintain the localized nature of the \textit{a priori}
constructed localized initial guess orbitals (\textit{vide infra}). Due to the
invariance of single determinants with respect to orbital rotations within the
occupied subspace the resulting minimizing single-determinant wave function is
equivalent with the usual canonical solution, which can be obtained from the
converged RSH matrix by a single diagonalization step.  

Having the localized RSH orbitals and the corresponding Fock matrix elements
at hand (see Section \ref{sec:orbitals}), the remaining long-range correlation
part in equation (\ref{eq:srlr}) should be calculated. In many-body
perturbational approaches special attention should be paid to the nonlinear
nature of the Hamiltonian, which may lead to additional contributions as
compared to conventional perturbation methods Ref.\ \cite{Angyan_deriv} (see
also \cite{Fromager:08}). 

The separation of the correlation energy to a short-range DFT part and a
long-range wave function part reduces significantly the dependence on basis
sets~\cite{Hertwig_basis_1995}, since the most strongly basis set dependent
electron-electron cusps are taken into account quite well by the density
functional part.  

Additionally, the size of the individual integrals shows a separation into
important contribution, arising from spatially close-lying orbitals, and less
important ones from distant molecular orbitals. These two aspects should
render the scheme in localized orbitals appealing.  

\subsection{The Random Phase Approximation} 

The Random Phase Approximation has become quite popular as a post-DFT
correlation method~\cite{eshuis_electron_2012}. It includes an infinite
summation of correlation diagrams, beyond second-order perturbation theory,
and it is invariant to orbital rotations\cite{bouman_optical_1979}  contrary
to, for instance, Epstein-Nesbet perturbation theory. It has been shown that
RPA is equivalent with a CCD (coupled cluster doubles) approximation~\cite{scuseria_ground_2008}. 
We use RPA as a long-range correlation method in the following.

Among the different ways of expressing the RPA correlation
energy~\cite{Angyan:11}, we have chosen the
adiabatic-connection-fluctuation-dissipation-theorem (ACFDT)
equation~\cite{ACFDT_2011}.  In this formulation the RPA correlation energy is
written as an integral over a coupling constant $\alpha$:

\begin{equation}
  E_c=\frac{1}{2} \int_0^1 d\alpha \; tr\left\{ \mathbb{F}^{lr} \mathbb{P}^{lr}_{c,\alpha} \right\}
  \label{eq:ACFDT}
\end{equation}

The coupling constant $\alpha$ scales the electron-electron interaction, and adiabatically
connects the physical system ($\alpha=1$) to the RSH reference system ($\alpha=0$). The correlation
energy $E_c$ is then logically written as the previous integral, between $\alpha=0$ and $\alpha=1$.

$\mathbb{F}^{lr}$ is a matrix involving the long-range two-electron integrals
and $\mathbb{P}^{lr}_{c,\alpha}$ is the correlation part of the two-particule density matrix, 
obtained from the solution vectors of the long-range RPA equations.
The size of all matrices is  given by the product of the number of occupied and virtual
orbitals ($n_{occ} \times n_{vir}$), \textit{i.e.} the number of single excitations.

Several versions of long-range RPA co-exist in the literature~\cite{Angyan:11,toulouse_closed-shell_2011}, depending whether exchange
is included in the kernel of the long-range RPA equations used to compute
$\mathbb{P}^{lr}_{c,\alpha}$ and whether anti-symmetrized integrals are used in the
definition of $\mathbb{F}^{lr}$. In the direct-RPA (dRPA)
version of long-range RPA one neglects exchange, while in the RPA-exchange (RPAx) variant
the exchange is included in the kernel. Further variants of these two versions
of long-range RPA can be sought depending on using or dropping antisymmetrized integrals
when forming the ACFDT integrand. The variants without anti-symmetrization are
labelled dRPA-I and RPAx-I (\textit{-I} for single-bar), whereas variants with
antisymmetrized integrals are named dRPA-II and RPAx-II (\textit{-II} for
double-bar). For a detailed overview on these energy expressions, see Ref.\
\cite{ACFDT_2011}.

For the present calculations, the version RPAx-I is
employed throughout, as this flavor of long-range RPA seems to yield the most reliable
results in a range-separated context~\cite{zhu_range-separated_2010}. In this version, we have :

\begin{equation}
  \mathbb{F}^{lr}=\left(\begin{array}{cc}
               \mathbf{K}&\quad\mathbf{K} \\
               \mathbf{K}&\quad\mathbf{K} 
             \end{array}\right)
\end{equation}
with $K_{ia,jb}=\langle ab | ij \rangle^{lr}$, and $\mathbb{P}^{lr}_{c,\alpha}$ is constructed from the solution vectors of :

\begin{equation}
 \left( \begin{array}{cc}
    \mathbf{\epsilon}+\alpha\mathbf{A}^{\prime}&\quad                  \alpha\mathbf{B}           \\
                      \alpha\mathbf{B}         &\quad\mathbf{\epsilon}+\alpha\mathbf{A}^{\prime}
 \end{array}\right)
 \left( \begin{array}{c}
    \mathbf{X}_{n,\alpha} \\
    \mathbf{Y}_{n,\alpha}
 \end{array}\right) = 
 \omega_{n,\alpha}
 \left( \begin{array}{c}
     \mathbf{X}_{n,\alpha} \\
    -\mathbf{Y}_{n,\alpha}
 \end{array}\right)
 \label{eq:Q}
\end{equation}
with $\mathbf{A}^{\prime}_{ia,jb} = \langle ib || aj \rangle^{lr}$, 
     $\mathbf{B}         _{ia,jb} = \langle ab || ij \rangle^{lr}$, and
     $\mathbf{\epsilon}  _{ia,jb} = F^{lr}_{ab}\delta_{ij}- F^{lr}_{ij}\delta_{ab}$, 
where $F^{lr}$ is the corresponding RSH operator.

\subsection{Localized orbitals}
\label{sec:orbitals}
Many methods are available to generate localized {\sl occupied} orbitals, by
projection or maximization of functionals like the common
Foster-Boys, Pipek-Mezey, von Niessen, or Edmiston-Ruedenberg methods. All these 
can be
applied equally well to Hartree-Fock or Kohn-Sham orbitals. {\sl Virtual}
localized orbitals may be generated as non-orthogonal projected atomic
orbitals~\cite{Pulay_localizability_1983}, as pair natural
orbitals~\cite{neese_efficient_2009-1}, as Optimized Virtual Orbital Space,
OVOS~\cite{pitonak_optimized_2006},
as complementary Boys-localized orbitals, or simply Pipek-Mezey localized
ones, occupying the least number of expansion centers per orbital.  

Having well-localized orbitals within the monomers is crucial to render the
selection of the excitations efficient, and to allow us to  separate inter-
and intra-molecular excitations in the context of intermolecular
interactions. To obtain this separation automatically, without the need for a
beforehand construction of canonical, delocalized molecular dimer orbitals, we
choose a slightly different route than the previously cited ones: localized
occupied and virtual orbitals will be constructed on the same footing. First
the RSH equations are solved in canonical orbitals for the monomers
in the monomer basis, and for the monomers in the dimer basis. Starting from
the first set, the orbitals are localized through a projection procedure,
described in reference~\cite{maynau_direct_2002}, and recalled in  
the appendix. This method constructs a minimal set of molecular occupied and
virtual orbitals, and adds the remaining virtual space as projected,
orthogonalized atomic orbitals.   

With these localized monomer orbitals in the {\sl monomer} basis and a converged
Kohn-Sham operator of the monomers in the {\sl dimer} basis, a Singles-CI procedure
is used to add the ghost basis set as supplementary virtual orbitals.
 
From the calculation of the monomers in the monomer basis the virtual orbitals
are taken, and from the monomers in the dimer basis the occupied ones, in
order to generate starting orbitals for the Singles-CI orbital optimization of
the real dimer system~\cite{reinhardt_fragment-localized_2008}. In this way the
orbitals and the corresponding excitations can be clearly identified in the
respective monomer and the dimer systems.   

At the end, three orbital sets are available (monomer orbitals in the monomer
basis, counterpoise-corrected monomer orbitals and dimer
orbitals), generated from one single guess and with overlaps in the vicinity
to one. 

\subsection{Selection of single excitations}
As it has been mentioned, the localized-orbitals framework can reduce the
computational effort by reducing the number of significant long-range two-electron
integrals to be taken into account.  
In order to reduce the dimensions of the matrices to be constructed and
treated for the long-range RPA calculations, further considerations are needed to
optimize the selection of determinants. A possibility is to use an energy
criterion of perturbative nature, by considering the second-order
approximation to long-range RPA, which happens to be the standard MP2 energy expression
in the RPAx-I variant. It will be supposed that 
only those single excited determinants $\Phi_i^a$ ({\it i.e.} a
single excitation),  which satisfy the following condition: 

\begin{equation}
\frac{\langle \Phi_0| \hat{{W}_{\text{ee}}^{\text{lr}}} | 
\Phi_{i\bar{i}}^{a\bar{a}}\rangle^2}{E(\Phi_{i\bar{i}}^{a\bar{a}})-E_0}\ = \ 
\frac{({\langle ii|aa \rangle}^{lr})^{2}}{2(\epsilon_a-\epsilon_i)} \ > \ \tau
\end{equation} 
are going to provide a significant contribution to the long-range RPA energy. This
evaluation is very rapid, since it implies only long-range two-electron integrals and
diagonal elements of the RSH matrix.
$|\Phi_0 \rangle$ and $E_0$ being respectively the RSH wavefunction and energy,
$\epsilon_a$ and $\epsilon_i$ repectively the RSH virtual and occupied orbital energies, 
 $\hat{{W}_{\text{ee}}^{\text{lr}}}$ the range separation operator from the equation \ref{eqn:rs},
$\Phi_{i\bar{i}}^{a\bar{a}}$ a determinant of two single excitations, and $E(\Phi_{i\bar{i}}^{a\bar{a}})$
its corresponding energy.
 The dimension of the long-range RPA matrix
is exactly the number of singly excited determinants thus selected, in
contrast to Configuration-Interaction-based methods, where the matrix
dimension is roughly proportional to the square of the selected
single excitations.   

Beyond the selection through the energetic importance of a determinant and in
the context of inter-molecular interactions single excitations can be classified
into intra-molecular and inter-molecular ones. The former ones have complete
analogs in the individual monomer system, as ideally orbitals are little
changed during the construction of the dimer wave function. The latter ones
should be of limited importance as the correlation part of the interaction
between monomers is mainly governed by dispersion-type interactions, which
imply two coupled intra-monomer excitations.

A significant saving for the calculation of the correlation contribution to
the interaction energy is obtained when (1) considering only single excitations
within the same monomers, assuming that inter-monomer excitations have only
small amplitudes with our choice of the construction of the orbitals perfectly
centered on the individual monomers, and (2) considering for the calculation
of the long-range RPA correlation contribution to the interaction energy only the
dispersion-type combination of the intra-monomer single excitations (see the
schema of figure \ref{figure1}). This assumes that the intra-monomer
correlation contributions are about the same  for the dimer and the isolated
monomers, and should not contribute significantly to the binding. The partial
summation over the long-range RPA amplitudes exploits the property that the correlation
energy is a linear functional of the amplitudes.

%For calculating these we need much less information than for the whole long-range RPA
%evaluation: the list of determinants does not include those with inter-monomer
%excitations, and only the evaluation of the long-range RPA equations for the dimer system
%is required (see the schematic figure \ref{figure1}).

%$$ {\rm ---- Fig.\ \ref{figure1}\  here\ ---- } $$
\bild{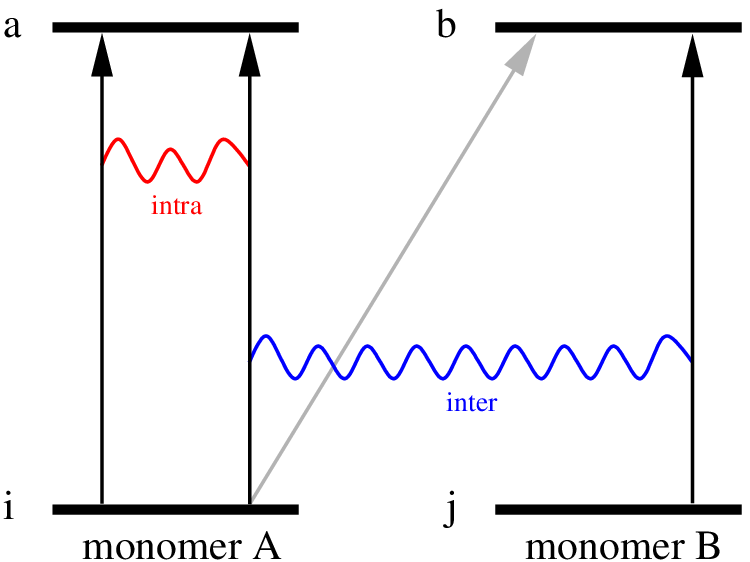}{8}{Single excitations (arrows) between an occupied ($i$ or
  $j$) and a virtual ($a$ or $b$) orbital in two monomers. These
  single excitations can either be intra-molecular (black arrows) or
  inter-molecular (gray arrow). One can then consider the coupling of two
  intra-molecular single excitations either within the same monomer (red double
  wavy line) or between two monomers (blue wavy line). These last couplings we
  call `dispersion-type combination of intra-monomer excitations'.}   

\section{Results}
\label{sec:results}

In srDFT+lrRPA, one can tweak the functional, the version of long-range RPA and the range
separation parameter to obtain a kind of best combination. The present study
aims, however, at providing insights in the general context of the
srDFT+lrRPA. We will fix the functional as the
srPBE~\cite{goll_short-range_2005} and the long-range RPA version as the RPAx-I (see
previous sections), and vary only the range-separation parameter. Conclusions
should not be affected by the specific choices made before.   

We will study a few selected dispersive and hydrogen-bonded systems:
the dimers of water (one hydrogen bond), methane (dispersion-only), formamide
(two hydrogen bonds) and, finally, the larger benzene dimer, bound by
dispersion interactions. Geometries are taken from the S22 test
set~\cite{jurecka_benchmark_2006}, and as basis set we use (excepted the
benzene dimer) the one given by Voisin~\cite{theseVoisin}, which has been
designed specifically for intermolecular interactions (see Appendix for more
details). For benzene we employed the standard aug-cc-pvdz
basis~\cite{Dunning_gaussian_1989}. All correlation energies are evaluated
with frozen core orbitals. 

If we look at the overlaps produced by the proposed Singles-CI procedure (see
section \ref{sec:orbitals}), we have more than 99.9$\,$\%\ throughout for the
{\sl occupied} dimer orbitals and the monomer orbitals, and 0.96$\pm$0.06,
0.97$\pm$0.05, 0.96$\pm$0.07 and 0.99$\pm$0.03, repectively, for the virtual
dimer and monomer orbitals for the four complexes of water, methane, formamide
and benzene. The procedure of perturbing the monomer orbitals through single
excitations toward the dimer orbitals deforms indeed only very slightly the
occupied orbitals, and little the virtual ones. 

\subsection{Correlation energy differences}
In this first part we employ the selection of single excitations individually
to the dimer and the separate monomers. Applying selection criteria of
$10^{-8}$, $10^{-9}$, and $10^{-10}\,$Hartrees, for a range separation
parameter $\mu=0.5\,$a.u. the results converge to the result obtained
without any selection, as displayed in Table \ref{tab:f12_threshold}, together
with reference data of explicitely correlated coupled-cluster (F12-CCSD(T))
results of Ref.~\cite{marchetti_accurate_2009}. Larger thresholds than
$10^{-8}\,$Hartree lead eventually to repulsive lr-correlation contributions.

%$$ {\rm ---- Table\ \ref{tab:f12_threshold}\  here\ ---- } $$
\begin{table}[!htb]
%\footnotesize
  \begin{center}
    \begin{tabular}{lcccc}
    $\Delta E$(mH)/dimer       & water &methane &formamide &benzene\\
    \hline
    srPBE                      & $-$7.423 &  +0.393  & $-$22.98 & $+$0.898 \\
    +lrRPAx-I ($10^{-8}$)      & $-$7.548 &  +0.307  & $-$24.21 & $-$0.420 \\
    +lrRPAx-I ($10^{-9}$)      & $-$8.178 & $-$0.140 & $-$25.44 & $-$1.742 \\
    +lrRPAx-I ($10^{-10}$)     & $-$8.396 & $-$0.500 & $-$25.91 & $-$2.880 \\
    +lrRPAx-I (all)            & $-$8.628 & $-$0.664 & $-$27.27 & $-$3.759 \\
    \hline                                                     
    F12-CCSD(T)    & $-$7.865$^a$ &
    $-$0.818$^a$ & $-$25.38$^a$ & $-$4.318$^a$ \\
    \hline
    \multispan5{$^a$from Ref.\
    \cite{marchetti_accurate_2008}, aug-cc-pvtz basis set\hfil}
  \end{tabular}
  \end{center}
  \caption{srPBE and srPBE+lrRPAx-I (for a range separation parameter of
    $\mu$=0.5$\,$a.u.) interaction energies for the water, methane 
    and formamide dimers in the Voisin-ANO basis and the benzene dimer in the
    aug-cc-pvdz basis, for different selection thresholds. All core orbitals
    are frozen, and results are BSSE corrected. For comparison we give as well
    published explicitly correlated coupled-cluster results.}  
  \label{tab:f12_threshold}
\end{table}

The possible savings while applying the selection criteria can be
seen in Figure \ref{figure2}, where we display the the evolution of the
individual correlation energies with the number of selected determinants. We
see that with roughly half of the determinants more than 90$\,$\% of 
the complete long-range RPA energy is recovered. Moreover, the selection criterion
becomes more and more efficient the larger the dimer systems is, as 
the overlaps between occupied and virtual orbitals become less and less
important. The figure gives as well an impression of the orders of magnitude
between the actual correlation energies and the contribution to the
interaction, leading to the slow convergence of the selection scheme as
applied here.

%$$ {\rm ---- Fig.\ \ref{figure2}\  here\ ---- } $$
\bild{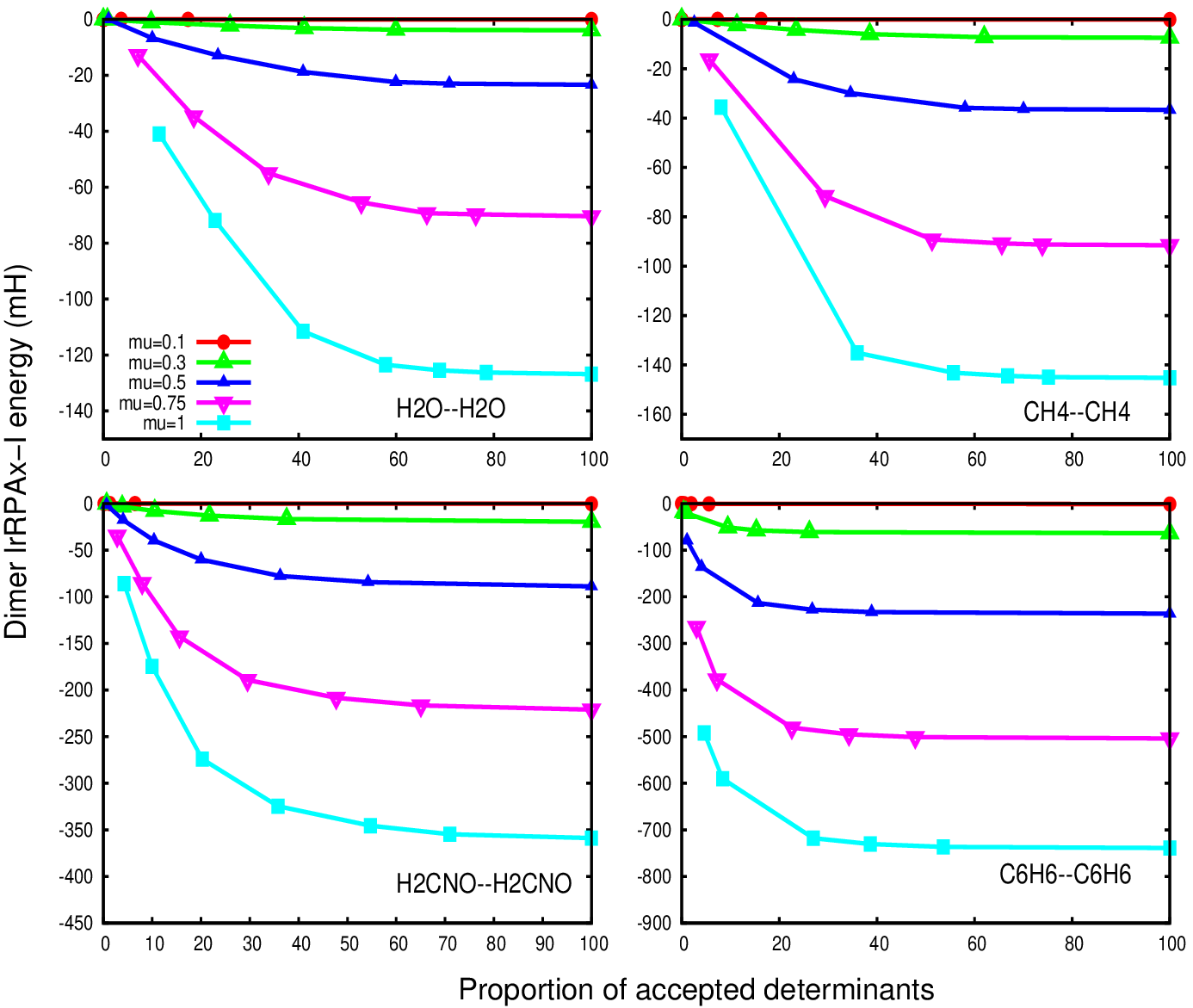}{15}{Evolution of the lrRPAx-I dimer energies (in mH) with the
  number of selected determinants (in \%\ of the possible determinants) for
  thresholds of $10^{-5}$, 10$^{-6}$, 10$^{-7}$, 10$^{-8}$, 10$^{-9}$,
  10$^{-10}$ Hartrees (from left to right, as tighter 
  thresholds select more determinants) for the water (top left), the methane
  (top right), the formamide (bottom left) and the benzene dimer (bottom
  right).}

\subsection{Dispersion-only contribution to the interaction energy}
As we see the poor convergence of the correlation contributions to the
interaction energy with tighter selection criteria, a scheme selecting
determinants homogeneously within the monomers and the dimer may be more
adequate, making use of the clear identifiability of the molecular orbitals of
monomers and the dimer. We propose here to regard only 
dispersion-type correlation diagrams (see figure \ref{figure1}) of 
all possible classes of diexcitations in the dimer system, without considering 
the monomers explicitly, but as furnishing starting orbitals for constructing
the monomer-localized dimer orbitals. 

As a supplementary simplification in terms of computational effort, we may
construct and solve the long-range RPA equations only within that space of single
excitations, neglecting all excitations having the occupied and virtual orbital
on different monomers. In addition, as in the previous section, the single
excitations can be selected through the previously employed scheme.

We present in Figure \ref{figure3}\ the result of this selection of
dispersion-only 
correlation contributions, for the commonly 
employed range-separation parameter $\mu=0.5\,$a.u. As the correlation energy
is displayed at a whole, all the inter-monomer excitations other than dispersion
are missing in the right-most bar of the
graph. Nevertheless, the dispersion part (topmost part of the bars), is nearly
unchanged between the two dimer calculations. As the attribution to the
different classes of diexcitations follows the attribution of orbitals to the
monomers, the explicit form of the starting monomer orbitals (canonical,
localized, approximated ...) has no impact on the final result. 

%$$ {\rm ---- Fig.\ \ref{figure3}\  here\ ---- } $$
\bild{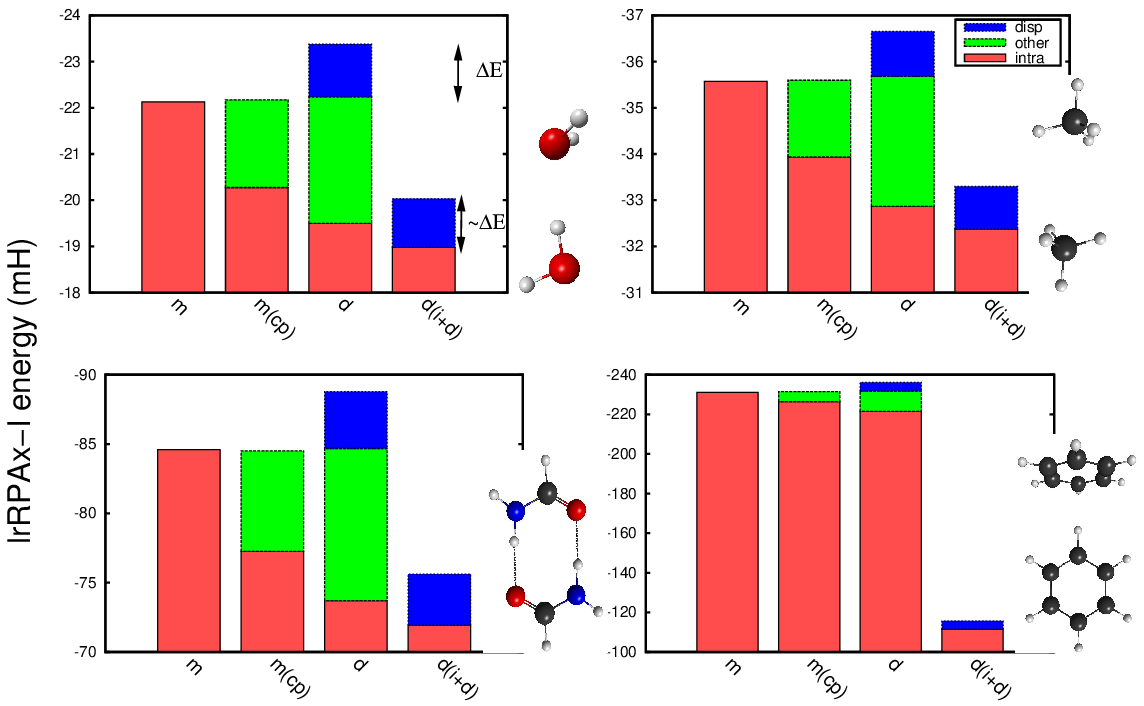}{11}{Decomposition of the lrRPAx-I correlation energies into
  different classes of excitations for the water(top left), methane (top
  right), formamide (bottom left) and benzene (bottom right) dimer, for a
  range-separation parameter of $\mu=0.5\,$a.u.. The left-most bar in each
  part shows the monomer in the monomer basis (by definition only
  intra-molecular correlation energy), the next the monomer in the whole dimer
  basis (separation into intra-molecular and other contributions like BSSE),
  then the dimer correlation energy with the monomer part, the dispersion part
  and the remaining contributions, and finally, the right-most bar, the dimer
  correlation energy when using only the intra-molecular and dispersion-type
  excitations in the RPA equations. Note that the zero of the energy scale is
  not included in the diagram.} 
%$$ {\rm ---- Fig.\ \ref{figure4}\  here\ ---- } $$
\bild{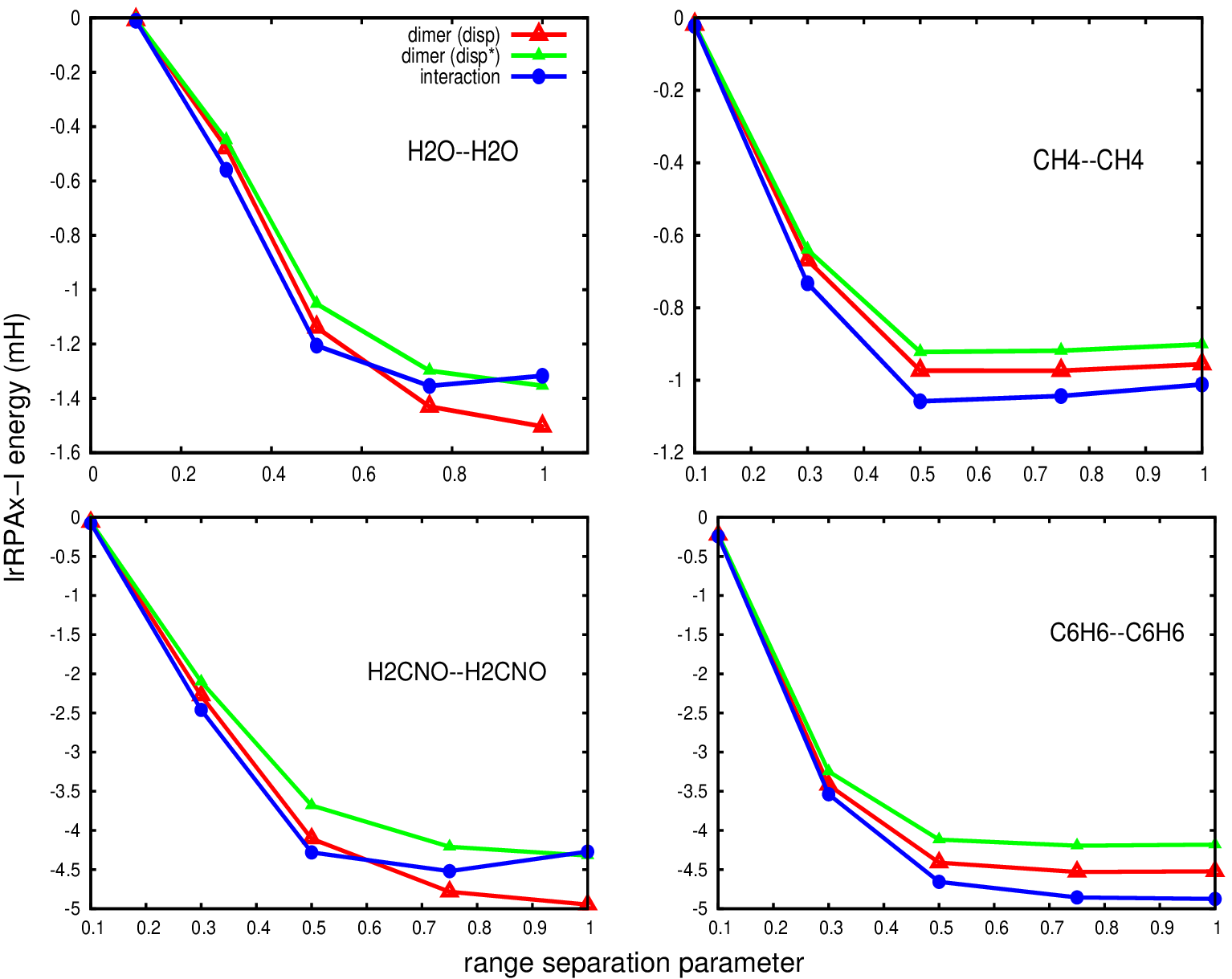}{11}{lrRPAx-I interaction energies in mH (in full circles) for the
  water(top left) , methane (top right), formamide (bottom left) and benzene
  (bottom right) dimer, versus the range separation parameter $\mu$. ``disp''
  (in hollow triangles) and ``disp*'' (in full triangles) represent the
  dispersion contribution to the dimer energy computed respectively from the
  whole RPA matrices and the intramolecular plus dispersion based RPA
  matrices.}

One should note the small difference of the monomer correlation
energies when including or not the ghost basis sets of the other monomer,
showing again the weak basis-set dependence of the RSH+lrRPA calculations. When
including the ghost basis sets, the excitations can be grouped into those
within the occupied and virtual orbitals of the monomer, and those from the
occupied orbitals toward the orbitals originating from the ghost basis
set. This part seems to be relatively large (green parts of the second monomer bar in
each panel), as the virtual space of the starting orbitals in this calculation
is the L\"owdin (or S$^{-1/2}$) orthogonalized ensemble of the virtual
orbitals 
of the monomer calculation in the monomer orbitals and the additional
ghost-basis atomic orbitals, without a hierarchy of virtual and ghost-orbital
space. As RPA is invariant to rotations of the orbital spaces, the overall
correlation energy is independent of this technical detail here, delocalizing
slightly orbitals due to the orthogonality constraints.      

We remark two coincidences that should not be fortuitous: the coincidence of 
a) the dispersion contribution with the contribution to the interaction energy on
 the one hand, and of b) the solution of the long-range RPA equation for all excitations
 with the solution of the long-range RPA equation only for the intra- and dispersion-type
 excitations on the other hand. 
Other correlation contributions to the
interaction energy like induction, due to the deformation of the monomers in
the dimer system and the corresponding change in correlation energy, are taken
into account by the short-range density functional. Only the dispersion should be
accounted for by the lr-RPA equations.

Changing $\mu$ from very small values (toward pure DFT calculations) and large
values (regular RPA) shows clearly this aspect (see Figure \ref{figure4}): the
long-range RPA contributions to the interaction energies deviates from the
dispersion-only approximation in the case of the two hydrogen-bound systems
for larger values of $\mu$. This qualitative difference in the weight of the
dispersion contributions to the long-range RPA interaction energies should be still more
pronounced in the case of charged monomers. This will be studied elsewhere.

As a final step we apply the selection of the single excitations to the
calculation of the dispersion-only correlation contribution to the interaction
energy. Table \ref{tab:f12_disp}\ shows the data for $\mu=0.5\,$a.u., as a
function of the selection threshold. 

%$$ {\rm ---- Table\ \ref{tab:f12_disp}\  here\ ---- } $$
\begin{table}[!htb]                                %\footnotesize
  \begin{center}
    \begin{tabular}{lcccc}
      E (mH)/dimer  & water & methane & formamide & benzene \\
      \hline
      $\Delta E_{srPBE}$   & $-$7.423 &  +0.393   & $-$22.984  &
      $-$0.564   \\ 
      &  &  &  & \\ 
      $+E_{lrRPAx-I}$     & $-$7.433 & +0.364 & $-$23.02  &  $-$0.074    \\ 
      (disp, $10^{-5}$)   & $-${\it 7.433}  &   {\it +0.364}  
      & $-${\it 23.02}  &  $-${\it 0.074}  \\ 
      $+E_{lrRPAx-I}$   & $-$7.758   &   +0.033  
      & $-$23.94   &  $-$1.101    \\ 
      (disp, $10^{-6}$)                            & $-${\it 7.758}  &   +{\it 0.033}  
      & $-${\it 23.94}  &  $-${\it 1.101}  \\ 
      $+E_{lrRPAx-I}$    & $-$8.021   & $-$0.294
      & $-$24.90   &  $-$2.301    \\ 
      (disp, $10^{-7}$)                       & $-${\it 8.031}  &
      $-${\it 0.296}   
      & $-${\it 24.98}  &  $-${\it 2.230}  \\ 
      
      $+E_{lrRPAx-I}$    & $-$8.299   & $-$0.398
      & $-$25.74   &  $-$2.943    \\ 
      (disp, $10^{-8}$)                       & $-${\it 8.331}  & $-${\it 0.407}
      & $-${\it 25.89}  &  $-${\it 2.806}  \\ 
      
      $+E_{lrRPAx-I}$    & $-$8.452   & $-$0.522
      & $-$26.37   &  $-$3.300   \\ 
      (disp, $10^{-9}$)                       & $-${\it 8.510}  & $-${\it 0.554}  
      & $-${\it 26.62}  &  $-${\it 3.088}  \\
      
      $+E_{lrRPAx-I}$   & $-$8.474  & $-$0.528
      & $-$26.57   &  $-$3.429    \\ 
      (disp, $10^{-10}$)                       & $-${\it 8.547}  & $-${\it 0.568}  
      & $-${\it 26.89}  &  $-${\it 3.180}  \\ 
      
      $+E_{lrRPAx-I}$            & $-$8.475  & $-$0.528
      & $-$26.67   &  $-$3.513    \\ 
      (disp)                         & $-${\it 8.562}  & $-${\it 0.579}  
      & $-${\it 27.09}  &  $-${\it 3.218}  \\ 
      & & & & \\ 
      +$\Delta E_{lrRPAx-I}$ (all)         & $-$8.628 & $-$0.664   & $-$27.27  &
      $-$3.759 \\ 
                                %\hline
      $\Delta E_{F12-CCSD(T)}$ & $-$7.865 &  $-$0.818  & $-$25.38  & $-$4.318          \\ 
    \end{tabular}
  \end{center}
  \caption{Interaction energies (in mH) for the water, methane, formamide and
    benzene dimer, for a range separation parameter of 0.5$\,$a.u., and
    different selection thresholds. The correlation contribution is evaluated
    from the dispersion-type part of the RPA energies, as part of the all dimer
    RPA components, and, in {\sl italics, from the solution of the RPA equations
      in intra- and dispersion parts only}. All core orbitals are frozen.}
  \label{tab:f12_disp}
\end{table}

The convergence is much more rapid than for the energy differences, however
not really smooth, as for instance the energy difference between selection
thresholds of 10$^{-7}$ and 10$^{-8}\,$Hartree are in the same order of
magnitude as for the difference of 10$^{-9}$ and 10$^{-10}\,$Hartree. We could
not yet determine a consistent extrapolation scheme to the final value, taking
into account all determinants without selection. Nevertheless, we see from the
table that taking into account all of the long-range RPA equations or just those for
intra- and dispersion-type diexcitations does not make a large difference. On the
contrary: the latter values are slightly closer to the complete long-range RPA
interaction without selection than the result of the evaluation of all of 
the long-range RPA equations. 

In terms of considered determinants, and thus the computational effort, we   
can go back to Figure \ref{figure2}, from which we see that for a threshold
of 10$^{-8}\,$Hartree only 40$\,$\%\ of the total number of determinants are
involved (for benzene only 20$\,$\%). This savings enter quadratically in the
number of matrix elements.  

\section{Conclusions}
In this article we propose to select excitations within long-range RPA
correlation corrections to range-separated hybrid density-functional theory. 

From the presented data we conclude that the chosen selection of excitations
via an energy criterion leads to imprecise results when applied separately to
the dimer system and the individual monomers, due to the relative smallness of
the calculated interaction energy (several percent of the individual
contributions only). In the context of range-separated density-functional
theory, however, we observe that the direct calculation of the long-range RPA
contribution to the interaction energy via the dispersion-type excitations in
the long-range RPA equations leads to about the same result as the complete long-range RPA
calculations without selections, with a much more rapid convergence towards the complete long-range
RPA interaction energy. The computational effort is significantly reduced as well, as a result of a much
smaller number of excitations involved and the unnecessary
evaluation of the counterpoise corrected monomers energy. This
seems consistently satisfied for the dispersion-type interactions of the
benzene and the methane dimer, and as well for the single and double hydrogen
bonds in the water and formamide dimer. For the latter, the good coincidence
of the dispersion-only and the complete long-range RPA calculation is lost for large
values of the range-separation parameter, showing that in these cases other
contributions than dispersion are not any more taken into account by the
short-range DFT functional.    

The construction of the orbital space (occupied and virtual), assigned to 
the monomers in a fragment-oriented approach is essential for this 
decomposition of the correlation energy. 

\section*{Acknowledgements}
All calculations have been carried out in the Laboratory of Theoretical
Chemistry in Paris. Support from the ANR project WADEMECOM was very helpful,
including the local version of the Molpro program package, developed within
this collaboration. Discussions with J.\ Toulouse, K.\ Sharkas and A.\ Savin
(Paris), and N.\ Ben Amor and D.\ Maynau (Toulouse) are gratefully
acknowledged. J.~G.~A.\ is grateful to the CEU IAS for the senior 
fellowship.

\section*{Technical details}

\subsection{The Voisin-ANO basis}
The ``Voisin basis'' is a 7s4p(O, N, C)/3s(H) type Van
Duijneveldt~\cite{van_duijneveldt_ibm_1971} basis, contracted from a 12s7p/6s
primitive basis, and augmented by
Voisin~\cite{Voisin_computation_1992}\ with diffuse and polarization functions, leading to a
13s8p3d/10s2p primitive basis, contracted once for each angular momentum to
lead to an overall basis set described as 8s5p3d/4s2p~\cite{Voisin_computation_1992}. This basis
set has been used in our group for previous  
studies~\cite{langlet_perturbational_1995,langlet_decomposition_2004}
on similar molecules.

The localization method~\cite{maynau_direct_2002} we use in this study
constructs in a first step local guess orbitals from linear combinations of
atom-like, step-wise (core, valence, ...) orthogonalized orbitals from a
minimal set of basis functions through the chemical intuition of the bonding
in a molecule  ($\sigma$ or $\pi$ bonding/antibonding orbitals, lone
pairs). Occupied orbitals from this construction should represent already
closely the electronic SCF density of the molecule, hence the use of an
Atomic Natural Orbital (ANO~\cite{foster_natural_1980,widmark_density_1990})  
basis set when possible. The guess orbitals are projected onto the occupied
space, then hierarchially orthogonalized, i.e.\ occupied orbitals among
themselves, then the occupied with the virtual orbitals and finally the
virtual and diffuse orbitals among themselves. 
 
From the completely decontracted Voisin basis we constructed thus 1$s$, 2$s$
and 2$p$ atomic orbitals with a modified atomic Hartree-Fock
program~\cite{roos_general_1968}, and we made sure that both the original and 
the partial-ANO Voisin basis yield the same RHF interaction energy for small
complexes. 

\subsection{Employed computer codes}
For the calculation of short-range DFT energies and the construction of the
RSH matrix in a given set of orbitals a development version of 
the Molpro package based on the 2010.1 release~\cite{Molpro} was
used. The SCF procedure itself and the hierarchical generation of orbital sets
(see section \ref{sec:orbitals}) was carried out with the local-orbital code
of Paris~\cite{ORTHO}. Long-range integrals for the long-range RPA part are calculated
using an intermediate version of the Dalton 2011 package~\cite{Dalton2011},
transformed on the molecular-orbital basis by a local program~\cite{ORTHO}, that was used as
well as for the selection of determinants and the generation of the input lists
for the evaluation of the long-range RPA energy. For that latter task routines have been
written (B.~M.) as part of a development version of Molpro and compiled as a
stand-alone tool reading the generated lists of integrals and excitations. 
In canonical orbitals, results are identical to corresponding calculations
employing Molpro.      

The explicit construction of starting orbitals relies on the code of the
university of Toulouse~\cite{maynau_direct_2002}.


\begin{thebibliography}{0}%
\makeatletter
\providecommand \@ifxundefined [1]{%
 \@ifx{#1\undefined}
}%
\providecommand \@ifnum [1]{%
 \ifnum #1\expandafter \@firstoftwo
 \else \expandafter \@secondoftwo
 \fi
}%
\providecommand \@ifx [1]{%
 \ifx #1\expandafter \@firstoftwo
 \else \expandafter \@secondoftwo
 \fi
}%
\providecommand \natexlab [1]{#1}%
\providecommand \enquote  [1]{``#1''}%
\providecommand \bibnamefont  [1]{#1}%
\providecommand \bibfnamefont [1]{#1}%
\providecommand \citenamefont [1]{#1}%
\providecommand \href@noop [0]{\@secondoftwo}%
\providecommand \href [0]{\begingroup \@sanitize@url \@href}%
\providecommand \@href[1]{\@@startlink{#1}\@@href}%
\providecommand \@@href[1]{\endgroup#1\@@endlink}%
\providecommand \@sanitize@url [0]{\catcode `\\12\catcode `\$12\catcode
  `\&12\catcode `\#12\catcode `\^12\catcode `\_12\catcode `\%12\relax}%
\providecommand \@@startlink[1]{}%
\providecommand \@@endlink[0]{}%
\providecommand \url  [0]{\begingroup\@sanitize@url \@url }%
\providecommand \@url [1]{\endgroup\@href {#1}{\urlprefix }}%
\providecommand \urlprefix  [0]{URL }%
\providecommand \Eprint [0]{\href }%
\providecommand \doibase [0]{http://dx.doi.org/}%
\providecommand \selectlanguage [0]{\@gobble}%
\providecommand \bibinfo  [0]{\@secondoftwo}%
\providecommand \bibfield  [0]{\@secondoftwo}%
\providecommand \translation [1]{[#1]}%
\providecommand \BibitemOpen [0]{}%
\providecommand \bibitemStop [0]{}%
\providecommand \bibitemNoStop [0]{.\EOS\space}%
\providecommand \EOS [0]{\spacefactor3000\relax}%
\providecommand \BibitemShut  [1]{\csname bibitem#1\endcsname}%
\let\auto@bib@innerbib\@empty
%</preamble>
\end{thebibliography}%


\begin{thebibliography}{50}
\expandafter\ifx\csname natexlab\endcsname\relax\def\natexlab#1{#1}\fi
\providecommand{\bibinfo}[2]{#2}
\ifx\xfnm\relax \def\xfnm[#1]{\unskip,\space#1}\fi
%Type = Article
\bibitem[{Patton and Pederson(1997)}]{patton_application_1997}
\bibinfo{author}{D.~C. Patton}, \bibinfo{author}{M.~R. Pederson},
  \bibinfo{journal}{Phys. Rev. A} \bibinfo{volume}{56} (\bibinfo{year}{1997})
  \bibinfo{pages}{2495--2498}.
%Type = Article
\bibitem[{Kristy{\'a}n and Pulay(1994)}]{Kristyan:94}
\bibinfo{author}{S.~Kristy{\'a}n}, \bibinfo{author}{P.~Pulay},
  \bibinfo{journal}{Chem. Phys. Lett.} \bibinfo{volume}{229}
  (\bibinfo{year}{1994}) \bibinfo{pages}{175--180}.
%Type = Article
\bibitem[{Harris(1985)}]{harris_simplified_1985}
\bibinfo{author}{J.~Harris}, \bibinfo{journal}{Phys. Rev. B}
  \bibinfo{volume}{31} (\bibinfo{year}{1985}) \bibinfo{pages}{1770--1779}.
%Type = Article
\bibitem[{Gerber and {\'A}ngy{\'a}n(2007)}]{Gerber:07}
\bibinfo{author}{I.~C. Gerber}, \bibinfo{author}{J.~G. {\'A}ngy{\'a}n},
  \bibinfo{journal}{J. Chem. Phys.} \bibinfo{volume}{126}
  (\bibinfo{year}{2007}) \bibinfo{pages}{044103}.
%Type = Inbook
\bibitem[{Koch and Holthausen(2001)}]{koch_guide_2001}
\bibinfo{author}{W.~Koch}, \bibinfo{author}{M.~C. Holthausen},
  \bibinfo{title}{A Chemist's Guide to Density Functional Theory},
  \bibinfo{publisher}{Wiley-VCH Verlag GmbH, Weinheim}, pp.
  \bibinfo{pages}{i--xiii}.
%Type = Article
\bibitem[{Wodrich et~al.(2006)Wodrich, Corminboeuf, and
  Schleyer}]{wodrich_systematic_2006}
\bibinfo{author}{M.~D. Wodrich}, \bibinfo{author}{C.~Corminboeuf},
  \bibinfo{author}{P.~v.~R. Schleyer}, \bibinfo{journal}{Organic Letters}
  \bibinfo{volume}{8} (\bibinfo{year}{2006}) \bibinfo{pages}{3631--3634}.
%Type = Article
\bibitem[{Andersson et~al.(1996)Andersson, Langreth, and
  Lundqvist}]{Langreth_1996}
\bibinfo{author}{Y.~Andersson}, \bibinfo{author}{D.~Langreth},
  \bibinfo{author}{B.~Lundqvist}, \bibinfo{journal}{Phys. Rev. Lett.}
  \bibinfo{volume}{76} (\bibinfo{year}{1996}) \bibinfo{pages}{102--105}.
%Type = Article
\bibitem[{Lee et~al.(2010)Lee, Murray, Kong, Lundqvist, and
  Langreth}]{Langreth_2010}
\bibinfo{author}{K.~Lee}, \bibinfo{author}{E.~Murray},
  \bibinfo{author}{L.~Kong}, \bibinfo{author}{B.~Lundqvist},
  \bibinfo{author}{D.~Langreth}, \bibinfo{journal}{Phys. Rev. B}
  \bibinfo{volume}{82} (\bibinfo{year}{2010}) \bibinfo{pages}{081101}.
%Type = Article
\bibitem[{Grimme(2004)}]{grimme_accurate_2004}
\bibinfo{author}{S.~Grimme}, \bibinfo{journal}{J. Comp. Chem.}
  \bibinfo{volume}{25} (\bibinfo{year}{2004}) \bibinfo{pages}{1463--1473}.
%Type = Article
\bibitem[{Grimme et~al.(2010)Grimme, Antony, Ehrlich, and Krieg}]{Grimme:10}
\bibinfo{author}{S.~Grimme}, \bibinfo{author}{J.~Antony},
  \bibinfo{author}{S.~Ehrlich}, \bibinfo{author}{H.~Krieg},
  \bibinfo{journal}{J. Chem. Phys.} \bibinfo{volume}{132}
  (\bibinfo{year}{2010}) \bibinfo{pages}{154104}.
%Type = Article
\bibitem[{Becke and Johnson(2005)}]{becke_exchange-hole_2005}
\bibinfo{author}{A.~D. Becke}, \bibinfo{author}{E.~R. Johnson},
  \bibinfo{journal}{J. Chem. Phys.} \bibinfo{volume}{122}
  (\bibinfo{year}{2005}) \bibinfo{pages}{154104}.
%Type = Article
\bibitem[{Tkatchenko and Scheffler(2009)}]{Tkatchenko:09}
\bibinfo{author}{A.~Tkatchenko}, \bibinfo{author}{M.~Scheffler},
  \bibinfo{journal}{Phys. Rev. Lett.} \bibinfo{volume}{102}
  (\bibinfo{year}{2009}) \bibinfo{pages}{073005}.
%Type = Incollection
\bibitem[{Savin(1996)}]{Seminario}
\bibinfo{author}{A.~Savin}, in: \bibinfo{editor}{J.~M. Seminario} (Ed.),
  \bibinfo{booktitle}{Recent developments and Applications of Modern Density
  Functional Theory}, \bibinfo{publisher}{Elsevier},
  \bibinfo{address}{Amsterdam}, \bibinfo{year}{1996}, pp.
  \bibinfo{pages}{327--257}.
%Type = Article
\bibitem[{\'Angy\'an et~al.(2005)\'Angy\'an, Gerber, Savin, and
  Toulouse}]{angyan_van_2005}
\bibinfo{author}{J.~G. \'Angy\'an}, \bibinfo{author}{I.~C. Gerber},
  \bibinfo{author}{A.~Savin}, \bibinfo{author}{J.~Toulouse},
  \bibinfo{journal}{Phys. Rev. A} \bibinfo{volume}{72} (\bibinfo{year}{2005})
  \bibinfo{pages}{012510}.
%Type = Article
\bibitem[{Goll et~al.(2006)Goll, Werner, Stoll, Leininger, {Gori-Giorgi}, and
  Savin}]{goll_short-range_2006}
\bibinfo{author}{E.~Goll}, \bibinfo{author}{H.~J. Werner},
  \bibinfo{author}{H.~Stoll}, \bibinfo{author}{T.~Leininger},
  \bibinfo{author}{P.~{Gori-Giorgi}}, \bibinfo{author}{A.~Savin},
  \bibinfo{journal}{Chem. Phys.} \bibinfo{volume}{329} (\bibinfo{year}{2006})
  \bibinfo{pages}{276--282}.
%Type = Article
\bibitem[{Toulouse et~al.(2009)Toulouse, Gerber, Jansen, Savin, and
  {\'A}ngy{\'a}n}]{Toulouse:09}
\bibinfo{author}{J.~Toulouse}, \bibinfo{author}{I.~C. Gerber},
  \bibinfo{author}{G.~Jansen}, \bibinfo{author}{A.~Savin},
  \bibinfo{author}{J.~G. {\'A}ngy{\'a}n}, \bibinfo{journal}{Phys. Rev. Lett.}
  \bibinfo{volume}{102} (\bibinfo{year}{2009}) \bibinfo{pages}{096404}.
%Type = Article
\bibitem[{Janesko et~al.(2009)Janesko, Henderson, and
  Scuseria}]{janesko_long-range-corrected_2009}
\bibinfo{author}{B.~Janesko}, \bibinfo{author}{T.~Henderson},
  \bibinfo{author}{G.~Scuseria}, \bibinfo{journal}{J. Chem. Phys.}
  \bibinfo{volume}{131} (\bibinfo{year}{2009}) \bibinfo{pages}{034110}.
%Type = Article
\bibitem[{Zhu et~al.(2010)Zhu, Toulouse, Savin, and
  \'Angy\'an}]{zhu_range-separated_2010}
\bibinfo{author}{W.~Zhu}, \bibinfo{author}{J.~Toulouse},
  \bibinfo{author}{A.~Savin}, \bibinfo{author}{J.~\'Angy\'an},
  \bibinfo{journal}{J. Chem. Phys.} \bibinfo{volume}{132}
  (\bibinfo{year}{2010}) \bibinfo{pages}{244108}.
%Type = Article
\bibitem[{Toulouse et~al.(2011)Toulouse, Zhu, Savin, Jansen, and
  \'Angy\'an}]{toulouse_closed-shell_2011}
\bibinfo{author}{J.~Toulouse}, \bibinfo{author}{W.~Zhu},
  \bibinfo{author}{A.~Savin}, \bibinfo{author}{G.~Jansen},
  \bibinfo{author}{J.~\'Angy\'an}, \bibinfo{journal}{J. Chem. Phys.}
  \bibinfo{volume}{135} (\bibinfo{year}{2011}) \bibinfo{pages}{084119}.
%Type = Article
\bibitem[{{\'A}ngy{\'a}n et~al.(2011){\'A}ngy{\'a}n, Liu, Toulouse, and
  Jansen}]{Angyan:11}
\bibinfo{author}{J.~G. {\'A}ngy{\'a}n}, \bibinfo{author}{R.-F. Liu},
  \bibinfo{author}{J.~Toulouse}, \bibinfo{author}{G.~Jansen},
  \bibinfo{journal}{J. Chem. Theory Comput.} \bibinfo{volume}{7}
  (\bibinfo{year}{2011}) \bibinfo{pages}{3116--3130}.
%Type = Article
\bibitem[{Kapuy and Kozmutza(1991)}]{Kapuy:91}
\bibinfo{author}{E.~Kapuy}, \bibinfo{author}{C.~Kozmutza}, \bibinfo{journal}{J.
  Chem. Phys.} \bibinfo{volume}{94} (\bibinfo{year}{1991})
  \bibinfo{pages}{5565--5573}.
%Type = Article
\bibitem[{Pulay(1983)}]{Pulay_localizability_1983}
\bibinfo{author}{P.~Pulay}, \bibinfo{journal}{Chem. Phys. Lett.}
  \bibinfo{volume}{100} (\bibinfo{year}{1983}) \bibinfo{pages}{151--154}.
%Type = Article
\bibitem[{Sch\"utz et~al.(1998)Sch\"utz, Rauhut, and
  Werner}]{schutz_local_1998}
\bibinfo{author}{M.~Sch\"utz}, \bibinfo{author}{G.~Rauhut},
  \bibinfo{author}{H.~Werner}, \bibinfo{journal}{J. Phys. Chem. A}
  \bibinfo{volume}{102} (\bibinfo{year}{1998}) \bibinfo{pages}{5997--6003}.
%Type = Article
\bibitem[{Ben~Amor et~al.(2011)Ben~Amor, Bessac, Hoyau, and
  Maynau}]{BenAmor_direct_2011}
\bibinfo{author}{N.~Ben~Amor}, \bibinfo{author}{F.~Bessac},
  \bibinfo{author}{S.~Hoyau}, \bibinfo{author}{D.~Maynau}, \bibinfo{journal}{J.
  Chem. Phys.} \bibinfo{volume}{135} (\bibinfo{year}{2011})
  \bibinfo{pages}{014101}.
%Type = Article
\bibitem[{Reinhardt et~al.(2008)Reinhardt, Piquemal, and
  Savin}]{reinhardt_fragment-localized_2008}
\bibinfo{author}{P.~Reinhardt}, \bibinfo{author}{J.~P. Piquemal},
  \bibinfo{author}{A.~Savin}, \bibinfo{journal}{J. Chem. Theo. Comp.}
  \bibinfo{volume}{4} (\bibinfo{year}{2008}) \bibinfo{pages}{2020--2029}.
%Type = Article
\bibitem[{{\'A}ngy{\'a}n(2008)}]{Angyan_deriv}
\bibinfo{author}{J.~G. {\'A}ngy{\'a}n}, \bibinfo{journal}{Phys. Rev. A}
  \bibinfo{volume}{78} (\bibinfo{year}{2008}) \bibinfo{pages}{022510}.
%Type = Article
\bibitem[{Fromager and Jensen(2008)}]{Fromager:08}
\bibinfo{author}{E.~Fromager}, \bibinfo{author}{H.~J.~A. Jensen},
  \bibinfo{journal}{Phys Rev A} \bibinfo{volume}{78} (\bibinfo{year}{2008})
  \bibinfo{pages}{022504}.
%Type = Article
\bibitem[{Hertwig and Koch(1995)}]{Hertwig_basis_1995}
\bibinfo{author}{R.~H. Hertwig}, \bibinfo{author}{W.~Koch},
  \bibinfo{journal}{J. Comp. Chem.} \bibinfo{volume}{16} (\bibinfo{year}{1995})
  \bibinfo{pages}{576--585}.
%Type = Article
\bibitem[{Eshuis et~al.(2012)Eshuis, Bates, and Furche}]{eshuis_electron_2012}
\bibinfo{author}{H.~Eshuis}, \bibinfo{author}{J.~E. Bates},
  \bibinfo{author}{F.~Furche}, \bibinfo{journal}{Theor. Chem. Acc.}
  \bibinfo{volume}{131} (\bibinfo{year}{2012}) \bibinfo{pages}{1084}.
%Type = Article
\bibitem[{Bouman et~al.(1979)Bouman, Voigt, and Hansen}]{bouman_optical_1979}
\bibinfo{author}{T.~Bouman}, \bibinfo{author}{B.~Voigt},
  \bibinfo{author}{A.~Hansen}, \bibinfo{journal}{J. Am. Chem. Soc.}
  \bibinfo{volume}{101} (\bibinfo{year}{1979}) \bibinfo{pages}{550--558}.
%Type = Article
\bibitem[{\'Angy\'an et~al.(2011)\'Angy\'an, Liu, Toulouse, and
  Jansen}]{ACFDT_2011}
\bibinfo{author}{J.~G. \'Angy\'an}, \bibinfo{author}{R.-F. Liu},
  \bibinfo{author}{J.~Toulouse}, \bibinfo{author}{G.~Jansen},
  \bibinfo{journal}{J. Chem. Theo. Comp.} \bibinfo{volume}{7}
  (\bibinfo{year}{2011}) \bibinfo{pages}{3116--3130}.
%Type = Article
\bibitem[{Neese et~al.(2009)Neese, Wennmohs, and
  Hansen}]{neese_efficient_2009-1}
\bibinfo{author}{F.~Neese}, \bibinfo{author}{F.~Wennmohs},
  \bibinfo{author}{A.~Hansen}, \bibinfo{journal}{J. Chem. Phys.}
  \bibinfo{volume}{130} (\bibinfo{year}{2009}) \bibinfo{pages}{114108}.
%Type = Article
\bibitem[{Pitonak et~al.(2006)Pitonak, Holka, Neogrady, and
  Urban}]{pitonak_optimized_2006}
\bibinfo{author}{M.~Pitonak}, \bibinfo{author}{F.~Holka},
  \bibinfo{author}{P.~Neogrady}, \bibinfo{author}{M.~Urban},
  \bibinfo{journal}{J. Mol. Struc.: {THEOCHEM}} \bibinfo{volume}{768}
  (\bibinfo{year}{2006}) \bibinfo{pages}{79--89}.
%Type = Article
\bibitem[{Maynau et~al.(2002)Maynau, Evangelisti, Guih\'ery, Calzado, and
  Malrieu}]{maynau_direct_2002}
\bibinfo{author}{D.~Maynau}, \bibinfo{author}{S.~Evangelisti},
  \bibinfo{author}{N.~Guih\'ery}, \bibinfo{author}{C.~Calzado},
  \bibinfo{author}{J.~Malrieu}, \bibinfo{journal}{J. Chem. Phys.}
  \bibinfo{volume}{116} (\bibinfo{year}{2002}) \bibinfo{pages}{10060}.
%Type = Article
\bibitem[{Goll et~al.(2005)Goll, Werner, and Stoll}]{goll_short-range_2005}
\bibinfo{author}{E.~Goll}, \bibinfo{author}{H.~J. Werner},
  \bibinfo{author}{H.~Stoll}, \bibinfo{journal}{Phys. Chem. Chem. Phys.}
  \bibinfo{volume}{7} (\bibinfo{year}{2005}) \bibinfo{pages}{3917--3923}.
%Type = Article
\bibitem[{Jurecka et~al.(2006)Jurecka, Sponer, Cerny, and
  Hobza}]{jurecka_benchmark_2006}
\bibinfo{author}{P.~Jurecka}, \bibinfo{author}{J.~Sponer},
  \bibinfo{author}{J.~Cerny}, \bibinfo{author}{P.~Hobza},
  \bibinfo{journal}{Phys. Chem. Chem. Phys.} \bibinfo{volume}{8}
  (\bibinfo{year}{2006}) \bibinfo{pages}{1985--1993}.
%Type = Phdthesis
\bibitem[{Voisin(1991)}]{theseVoisin}
\bibinfo{author}{C.~Voisin}, \bibinfo{title}{Contributions to the computation
  of the induction term in intermolecular potentials for the modelisation of
  polypeptids}, Ph.D. thesis, Nancy University, France, \bibinfo{year}{1991}.
  \bibinfo{note}{In french}.
%Type = Article
\bibitem[{Dunning(1989)}]{Dunning_gaussian_1989}
\bibinfo{author}{T.~H. Dunning}, \bibinfo{journal}{J. Chem. Phys.}
  \bibinfo{volume}{90} (\bibinfo{year}{1989}) \bibinfo{pages}{1007--1023}.
%Type = Article
\bibitem[{Marchetti and Werner(2009)}]{marchetti_accurate_2009}
\bibinfo{author}{O.~Marchetti}, \bibinfo{author}{H.~Werner},
  \bibinfo{journal}{J. Phys. Chem. A} \bibinfo{volume}{113}
  (\bibinfo{year}{2009}) \bibinfo{pages}{11580--11585}.
%Type = Techreport
\bibitem[{Van~Duijneveldt(1971)}]{van_duijneveldt_ibm_1971}
\bibinfo{author}{F.~B. Van~Duijneveldt}, \bibinfo{title}{Gaussian Basis Sets
  for the Atoms H--Ne for Use in Molecular Calculations},
  \bibinfo{type}{Technical Report} \bibinfo{number}{RJ 945}, IBM, San Jos\'e,
  \bibinfo{year}{1971}.
%Type = Article
\bibitem[{Voisin et~al.(1992)Voisin, Cartier, and
  Rivail}]{Voisin_computation_1992}
\bibinfo{author}{C.~Voisin}, \bibinfo{author}{A.~Cartier},
  \bibinfo{author}{J.~L. Rivail}, \bibinfo{journal}{J. Phys. Chem.}
  \bibinfo{volume}{96} (\bibinfo{year}{1992}).
%Type = Article
\bibitem[{Langlet et~al.(1995)Langlet, Caillet, and
  Caffarel}]{langlet_perturbational_1995}
\bibinfo{author}{J.~Langlet}, \bibinfo{author}{J.~Caillet},
  \bibinfo{author}{M.~Caffarel}, \bibinfo{journal}{J. Chem. Phys.}
  \bibinfo{volume}{103} (\bibinfo{year}{1995}) \bibinfo{pages}{8043}.
%Type = Article
\bibitem[{Langlet et~al.(2004)Langlet, Berg\`es, and
  Reinhardt}]{langlet_decomposition_2004}
\bibinfo{author}{J.~Langlet}, \bibinfo{author}{J.~Berg\`es},
  \bibinfo{author}{P.~Reinhardt}, \bibinfo{journal}{J. Mol. Struc.: {THEOCHEM}}
  \bibinfo{volume}{685} (\bibinfo{year}{2004}) \bibinfo{pages}{43--56}.
%Type = Article
\bibitem[{Foster and Weinhold(1980)}]{foster_natural_1980}
\bibinfo{author}{J.~P. Foster}, \bibinfo{author}{F.~Weinhold},
  \bibinfo{journal}{J. Am. Chem. Soc.} \bibinfo{volume}{102}
  (\bibinfo{year}{1980}) \bibinfo{pages}{7211--7218}.
%Type = Article
\bibitem[{Widmark et~al.(1990)Widmark, Malmqvist, and
  Roos}]{widmark_density_1990}
\bibinfo{author}{P.~Widmark}, \bibinfo{author}{P.~Malmqvist},
  \bibinfo{author}{B.~Roos}, \bibinfo{journal}{Theor. Chem. Acc.}
  \bibinfo{volume}{77} (\bibinfo{year}{1990}) \bibinfo{pages}{291--306}.
%Type = Techreport
\bibitem[{Roos et~al.(1968)Roos, Salez, Veillard, and
  Clementi}]{roos_general_1968}
\bibinfo{author}{B.~Roos}, \bibinfo{author}{B.~Salez},
  \bibinfo{author}{A.~Veillard}, \bibinfo{author}{E.~Clementi},
  \bibinfo{title}{Atomic Hartree-Fock program}, \bibinfo{type}{Technical
  Report} \bibinfo{number}{RJ 815}, IBM, San Jos\'e, \bibinfo{year}{1968}.
  \bibinfo{note}{Modified by L. Gianola (1978), later by B.A. He\ss{} (1986)
  and P. Reinhardt (1995)}.
%Type = Misc
\bibitem[{Werner et~al.(2010)Werner, Knowles, Knizia, Manby, Sch\"utz
  et~al.}]{Molpro}
\bibinfo{author}{H.-J. Werner}, \bibinfo{author}{P.~Knowles},
  \bibinfo{author}{G.~Knizia}, \bibinfo{author}{F.~Manby},
  \bibinfo{author}{M.~Sch\"utz}, et~al., \bibinfo{title}{Molpro, version
  2010.1, a package of ab initio programs}, \bibinfo{year}{2010}.
  \bibinfo{note}{See {\tt http://www.molpro.net}}.
%Type = Misc
\bibitem[{Reinhardt(dots)}]{ORTHO}
\bibinfo{author}{P.~Reinhardt}, \bibinfo{title}{Ortho, series of ab-initio
  programs in localized orbitals}, \bibinfo{year}{1996 -- $\ldots$}.
  \bibinfo{note}{Unpublished}.
%Type = Misc
\bibitem[{Ruud et~al.(2011)Ruud, Helgaker, Olsen, J\o{}rgensen, Jensen
  et~al.}]{Dalton2011}
\bibinfo{author}{K.~Ruud}, \bibinfo{author}{T.~Helgaker},
  \bibinfo{author}{J.~Olsen}, \bibinfo{author}{P.~J\o{}rgensen},
  \bibinfo{author}{H.~J.~A. Jensen}, et~al., \bibinfo{title}{Dalton2011, a
  molecular electronic structure program, see {\tt
  http://www.daltonprogram.org}}, \bibinfo{year}{2011}.
%Type = Article
\bibitem[{Marchetti and Werner(2008)}]{marchetti_accurate_2008}
\bibinfo{author}{O.~Marchetti}, \bibinfo{author}{H.~Werner},
  \bibinfo{journal}{Phys. Chem. Chem. Phys.} \bibinfo{volume}{10}
  (\bibinfo{year}{2008}) \bibinfo{pages}{3400--3409}.
%Type = Article
\bibitem[{Scuseria, Henderson and Sorensen(2008)}]{scuseria_ground_2008}
\bibinfo{author}{G. E. ~Scuseria}, \bibinfo{author}{T. M. ~Henderson},
  \bibinfo{author}{D. C. ~Sorensen},
  \bibinfo{journal}{J. Chem. Phys.} \bibinfo{volume}{129}
  (\bibinfo{year}{2008}) \bibinfo{pages}{231101}.
\end{thebibliography}
\end{document}